\documentclass[conference,a4paper]{APSIPA2019}
\usepackage{multirow}
\usepackage{graphicx}
\usepackage{amsmath}
\usepackage[psamsfonts]{amssymb}
\usepackage{amsxtra}
\usepackage{threeparttable}
\usepackage{subfigure}
\usepackage{comment}
\usepackage{hyperref}
\usepackage{url}
\usepackage{cite}
\usepackage{enumerate}
\begin{document}

\title{Improving Automatic Jazz Melody Generation \\by Transfer Learning Techniques}

\author{%
\authorblockN{%
Hsiao-Tzu Hung\authorrefmark{1}\authorrefmark{2},
Chung-Yang Wang\authorrefmark{2}
Yi-Hsuan Yang\authorrefmark{2}\authorrefmark{3},
Hsin-Min Wang\authorrefmark{1}
}
\authorblockA{%
\authorrefmark{1}
Institute of Information Science, Academia Sinica, Taipei, Taiwan \\}
\authorblockA{%
\authorrefmark{2}
Taiwan AI Labs, Taipei, Taiwan\\}
\authorblockA{%
\authorrefmark{3}
Research Center for IT Innovation, Academia Sinica, Taipei, Taiwan\\
E-mail: \{fbiannahung,wangyygogo\}@gmail.com, yhyang@ailabs.tw, whm@iis.sinica.edu.tw
}
}

\maketitle
\thispagestyle{empty}

\begin{abstract}
In this paper, we tackle the problem of transfer learning for Jazz automatic generation. Jazz is one of representative types of music, but the lack of Jazz data in the MIDI format hinders the construction of a generative model for Jazz. Transfer learning is an approach aiming to solve the problem of data insufficiency, so as to transfer the common feature from one domain to another. 
In view of its success in other machine learning problems,
we investigate whether, and how much, it can help improve automatic music generation for under-resourced musical genres.  Specifically, we use a recurrent variational autoencoder as the generative model, and use a genre-unspecified dataset as the source dataset and a Jazz-only dataset  as the target dataset. Two transfer learning methods are evaluated using six levels of source-to-target data ratios. The first method is to train the model on the source dataset, and then fine-tune the resulting model parameters on the target dataset. The second method is to train the model on both the source and target datasets at the same time, but add genre labels to the latent vectors and use a genre classifier to improve Jazz generation. The evaluation results show that the second method seems to perform better overall, but it cannot take full advantage of the genre-unspecified dataset.
\end{abstract}

\section{Introduction}
Deep learning-based machine learning algorithms have been increasingly  employed for automatic music composition in recent years \cite{midinet,melodyRNN16,Jazzgan}. 
Typically, this is done by collecting a large dataset of machine-readable musical scores of existing music, in formats such as MIDI files,\footnote{[Online] \url{https://www.midi.org/}} and then using neural network models, such as the generative adversarial model (GAN) \cite{gan} and  variational autoencoder (VAE) \cite{vae}, to learn to compose new music via learning from the provided dataset. For example, the Lakh Pianoroll Dataset (LPD) is a public-domain dataset compiled by Dong \emph{et al.} for building a GAN model for multitrack music composition \cite{musegan}; it encompasses 50,266 four-bar MIDI phrases of Rock/Pop music in 4/4 time signature. 
As another example, Roberts \emph{et al.} \cite{musicvae} attempted to collect a large number of MIDI files from the Web to train a VAE model for generating melodies (monophonic note sequences) for unspecified musical genres; it is said that around 1.5 million unique MIDI files were found and downloaded.

The main advantage of such deep learning-based models for automatic music composition, compared to the rule-based or genetic algorithm-based algorithms studied by researchers decades ago \cite{fernandez13jair}, appears to be the unprecedented ability of deep learning models to find their own ways in learning from big data. We have already seen from the literature promising examples that use deep learning to learn to compose music for musical genres such as Rock \cite{musegan}, Pop \cite{song_from_pi}, and Classical music \cite{deepbach}.  
Common to these genres is the availability of MIDI files from the Web, providing sufficient data to train deep learning models.

\begin{table}
\caption{The percentage of melody labeled with different genre tags in the TheoryTab (TT) dataset. It contains 11,329 melodies, and is used as the ``source domain'' dataset in this work.}

\begin{tabular}{|ccccccc|}
\hline
Jazz   & Folk   & Dance  & Electronic & Rock   & Pop     & Unlabeled \\ \hline
2.12\% & 2.12\% & 6.23\% & 10.70\%    & 9.04\% & 11.25\% & 58.54\%   \\ \hline
\end{tabular}
\label{table:tt_genre_distribution}
\end{table}

However, this is not the case with many main musical genres in the world. An obvious example is Jazz, which often features live improvisations (i.e., with spontaneously invented melodic solo lines or accompaniment parts). In other words, a complete Jazz music piece is rarely composed \emph{offline} with a MIDI editor; rather, Jazz is usually created \emph{online} with spontaneous interaction among musicians. Extra effort is required to listen to the audio recording of a Jazz performance and carefully transcribe it by hand into a MIDI file. Consequently, MIDI files for Jazz music are relatively scarce on the Web. 

To illustrate this, we wrote a crawler to download a total of 11,329 melody phrases from an online music theory forum called TheoryTab.\footnote{TheoryTab is hosted by Hooktheory, a company that produces educational music software and books ([Online] \url{https://www.hooktheory.com/theorytab}).}
As shown in Table \ref{table:tt_genre_distribution}, only 2.12\% of the melodies were labeled as Jazz by the contributing forum users. Pop and Rock, for example, have around five times more data.

To our best knowledge, there is relatively little work on building automatic music composition models for Jazz. One prominent prior work is the work by Trieu and Keller \cite{Jazzgan}, who employed GAN to build a model called JazzGAN for chord-conditioned melody composition. However, likely due to the reasons outlined above, the dataset they used to train JazzGAN contained only 44 leadsheets, approximately 1,700 bars. 


In the machine learning community, many ``transfer learning'' techniques have been proposed to address the data scarcity of target tasks  \cite{transfer_survey,oquab,imagenet,bert}. The idea is to find a related \emph{source task} where the training data is easier to collect, and then adapt the model of the source task to the model of the \emph{target task} with a small dataset in the target domain. Given the relative richness of non-Jazz MIDI data, a natural research question is whether, and how much, we can leverage a large genre-unspecified \emph{source domain} MIDI dataset to improve the model for Jazz with a small genre-specific \emph{target domain} MIDI dataset.

In this paper, we aim to address the following research questions.
\begin{itemize}
    \item Can we use a genre-unspecified music dataset to improve a Jazz melody generation model?
    \item Which transfer learning technique is more useful for this task? 
    \item Does a transfer learning method benefit from increasing the size of the source domain data?
\end{itemize}
For the second research question, we evaluate two canonical transfer learning methods in this work: model fine-tuning and multitask learning (see Section \ref{section:method}). 
For the third research question, we consider six levels of source-to-target data ratios (see Section \ref{section:exp_baseline}).
For performance evaluation, we follow the recent work of Yang and Lerch \cite{yanglerch} and adopt seven different criteria for quantitative evaluation (see Section \ref{section:exp_metrics}).




While a piece of Jazz music can be composed of multiple tracks/instruments, we only focus on the melody in this work. In addition, we consider the task of ``unconditioned'' Jazz melody generation, i.e., generating melodies without any pre-determined conditions or information. This scenario is more challenging, yet practically more flexible and potentially more useful, than the ``chord-conditioned'' scenario addressed by JazzGAN \cite{Jazzgan}, where a sequence of accompanying chord labels is given to inform the melody generation model.

Certainly, Jazz is not the only ``under-resourced'' \cite{taslp17} genre in music. Since our problem formulation is generic, it is hoped that the lessons learned here can also be applied to other 
musical genres. In addition, the problem may be interesting not only for music AI researchers but also for general machine learning researchers, as transfer learning is more commonly employed for discriminative tasks such as classification and regression, rather than generative tasks such as automatic generation.

The 
paper is organized as follows. Section \ref{section:bg} provides background knowledge on transfer learning. Sections \ref{section:db} and \ref{section:method} present the datasets and models used in this work. Section \ref{section:exp} details on the evaluation setup. Section \ref{section:result} discusses the evaluation results.
Finally, Section \ref{section:conclusion} concludes the paper.

\section{Background}
\label{section:bg}
\subsection{Transfer Learning}
The general idea of transfer learning is to learn knowledge from one task and apply it to another task.  Generally, there will be two tasks A and B. Both tasks have the same input type, such as image and audio. Our main task is B, but the dataset for task B is much more smaller than the dataset for A, which may not be enough for training. Assuming that tasks A and B share some low-level features, we can improve task B by learning task A first. There are two training steps for transfer learning. The first step is to train the model on the dataset of task A, which is often referred to as ``pre-training." The second step is to further train the model obtained in the first step on the dataset of task B, which is called `fine-tuning." There are many transfer learning methods based on such a pre-training/fine-tuning strategy. An overview of recent techniques can be found in \cite{hylee}.

In recent years, transfer learning based on the above strategy has been widely used in many machine learning problems in computer vision (CV) and natural language processing (NLP). Well-known examples include the use of the first few layers of deep models trained on the ImageNet object recognition task as  visual feature extractors for other CV tasks \cite{imagenet} and the use of Google's pre-trained BERT model to get word and sentence embedding features for downstream NLP tasks \cite{bert}.

\subsection{Transfer Learning in Music-related Tasks}
Transfer learning techniques have also been applied to several discrminative tasks 
in the field of music information retrieval (MIR). For example, in \cite{choi17ismir}, a convnet was trained for music tagging, and then transferred to other music-related classification and regression tasks. The tags of the source task include genres, instruments, moods, and eras. The target tasks include vocal/non-vocal classification and general audio event classification. There are many other examples, all of which are concerned with classification or regression tasks \cite{davies13ismir,park18ismir,lu18ismir,luo18ismir}.

To our best knowledge, transfer learning techniques have not been used for automatic music composition. Researchers either work on genres with easy-to-access MIDI data (e.g., Rock and Pop) \cite{musegan} or a general model using genre-unspecified MIDI data \cite{musicvae}. 


\begin{figure*}[!t]
\centering
\includegraphics[width=0.6\columnwidth]{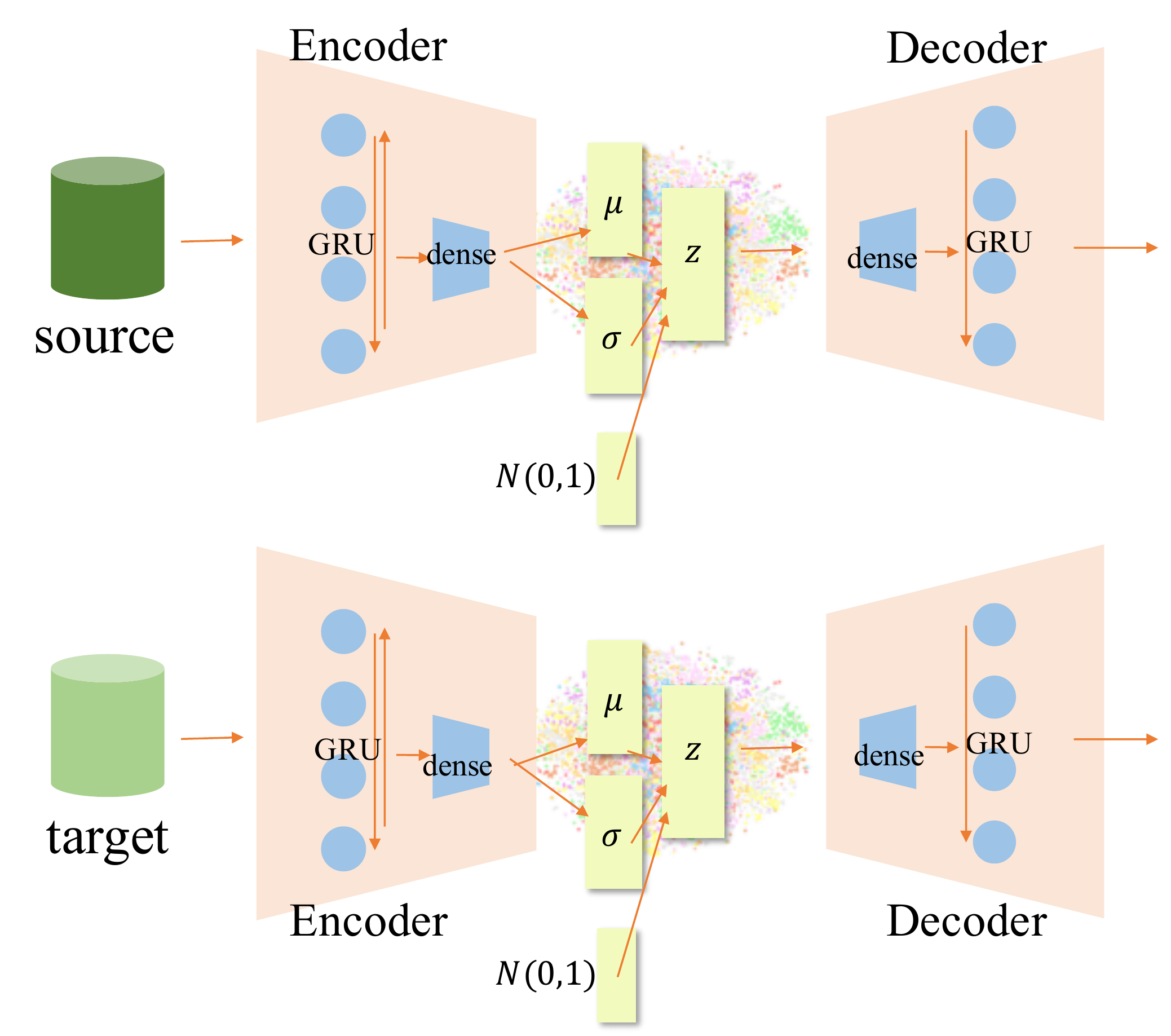}~~
\hspace{1cm}
\includegraphics[width=0.6\columnwidth]{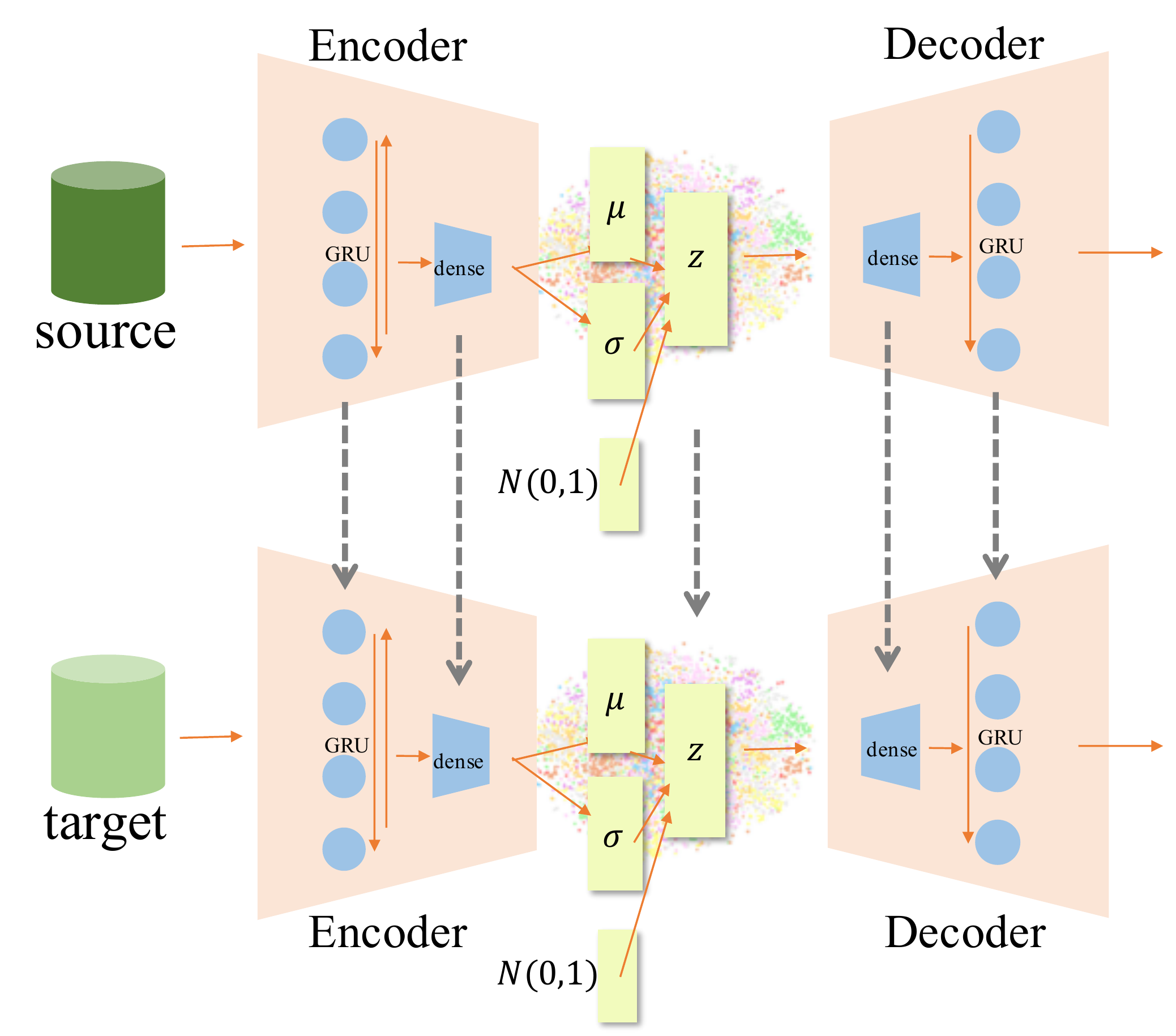}~~
\hspace{1cm}
\includegraphics[width=0.6\columnwidth]{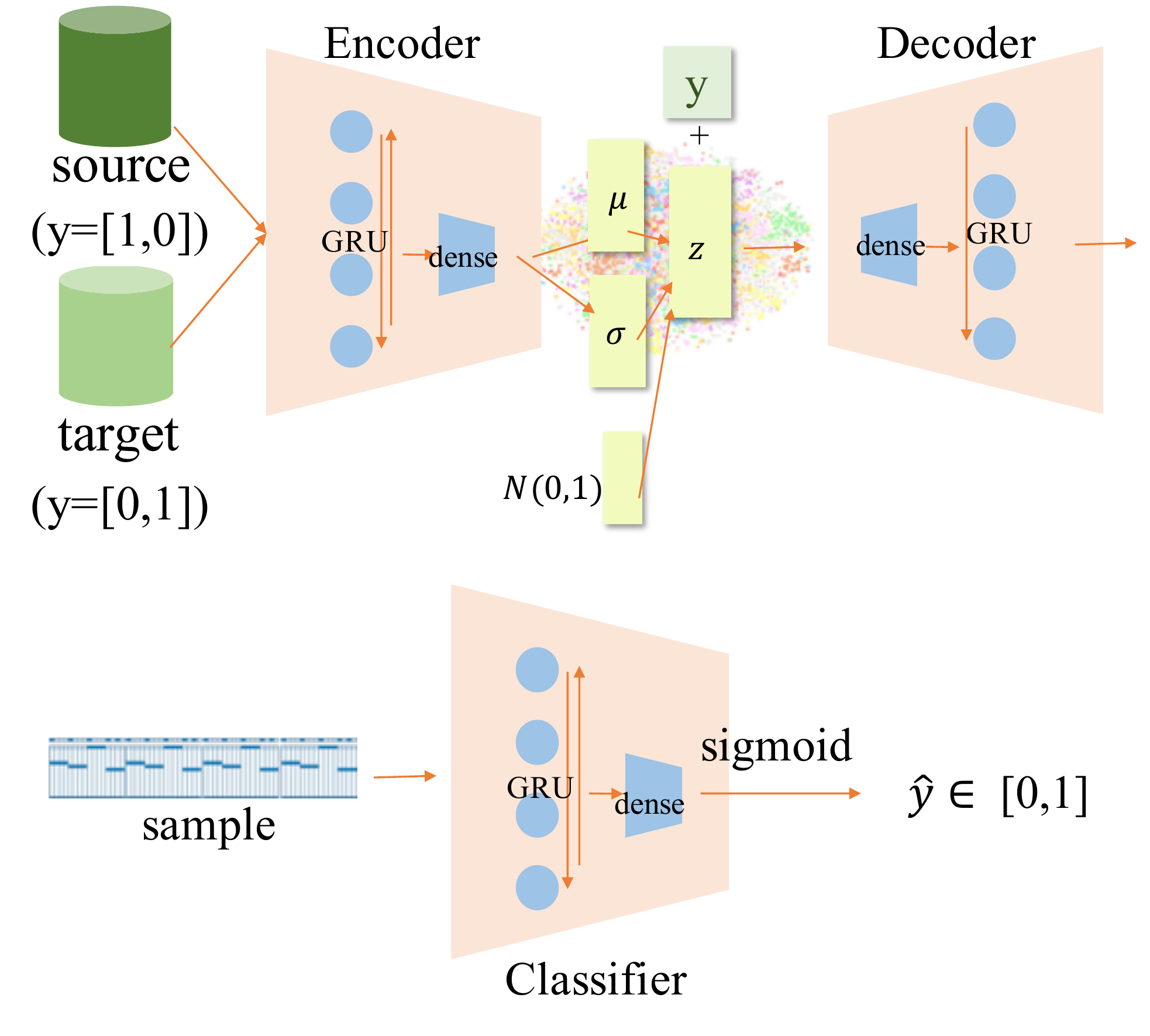}
\\ 
\hspace{0.6cm}(a)\hspace{6cm}(b)\hspace{6cm}(c)
\caption{Model architectures and training flows of the methods evaluated in this paper. \texttt{(a)} The baseline methods trained only on the source dataset (upper) or the target dataset (lower). \texttt{(b)} The ``fine-tuning'' method that pre-trains a model using the source dataset (upper) and then fine-tunes the model using the target dataset (lower). \texttt{(c)} The ``multi-task'' method that introduces an additional genre label $y$ as a conditional variable so as to train the model together on the source and target datasets (upper), with a separately trained genre classifier for further improving Jazz generation (lower).}
\label{fig:PP}
\end{figure*}

\begin{table}[]
\centering
\caption{The two datasets used in this work. A phrase is defined as a four-bar segment sampled from a song.}
\begin{tabular}{|l|rr|}
\hline
                & TT (source)        & CY$+$R (target)    \\ \hline
Genre           & diverse       & Jazz only \\
Song length     & segment       & segment   \\
Track           & melody, chord & melody    \\
Musical key     & C major, C minor         & C major \\
Time signature  & 4/4           & 4/4       \\
Number of phrases &9,640           &1,608        \\
Number of bars  & 38,560         & 6,432      \\ \hline
\end{tabular}
\label{table:tt_db}
\end{table}

\section{Datasets}
\label{section:db}

For this study, we have collected a clean Jazz-only dataset as the target dataset, and a genre-unspecified dataset as the source dataset.
The two datasets are summarized in Table \ref{table:tt_db}. 

The target dataset, referred to as the CY$+$R dataset hereafter, consists of two small Jazz music collections. The first collection consists of 575 four-bar melody phrases composed by one of the authors, who is a well-trained musician. All phrases are soft Jazz music. The second collection comes from the Jazz Realbook,\footnote{[Online]\url{https://www.profesordepiano.com/Real\%20Book/Realbook.htm}} which contains 240 unique songs. 

The musician manually split out a few four-bar phrases from each song, ensuring that each phrase has a musically plausible ending.
Finally, there are 1,608 four-bar phrases in CY$+$R, 1,446 phrases for training, and 162 phrases for testing.

The source dataset contains 11,329 melody phrases downloaded from TheoryTab\footnote{Note that the data scraped from TheoryTab is in the so-called lead sheet format, containing both the melody track and the chord track. We ignore the chord track in this work, since we consider unconditioned melody generation.}. As shown in Table \ref{table:tt_genre_distribution}, more than 50\% of the data are genre-unspecified. Although a small portion of them were labeled as Jazz, we checked the songs and found that some of the labels were unreliable. For example, ``Hard To Say I'm Sorry'' by Chicago was wrongly labeled as Jazz. In addition, most Jazz songs were style-wise quite different from our target dataset. Therefore, we ignored the genre labels and treated all the melodies as genre-unspecified. We use `TT' as the abbreviation for the TheoryTab dataset. Each song in TT is saved separately, including intro, chorus, verse, and outro. The intros of most songs in TT are composed of broken chords, which means repeating some notes from the chords, and they may not be considered as melodies. Therefore, the intros are excluded from the TT dataset, leaving 9,640 four-bar phrases. The ratio of training data to testing data is 9:1, i.e., 90\% for training, and 10\% for testing.

Table \ref{table:tt_db} summarizes the two datatsets. Both datasets are in 4/4 time signature. The CY+R dataset is written in C-major scale, and each melody phrase is transposed to the C major or C minor key in TT dataset. In order to broaden the diversity of generated melodies, we keep both C major and C minor songs in the TT dataset. 

\subsection{Representation of a melody phrase}
There are multiple ways to computationally represent a melody phrase. For example, Trieu and Keller investigated and compared the so-called ``event-based'' and ``time step-based'' representations of melody \cite{Jazzgan}. 
In this work, we adopt the time step-based method and represent each four-bar melody phrase as a fixed-size matrix.\footnote{One advantage of the time step-based representation over the event-based representation is that we can more easily emphasize the beat position of notes.}
This matrix-like representation has also been referred to as the \emph{pianoroll} representation \cite{pypianoroll}, where the horizontal axis denotes time (time step), and the vertical axis denotes frequency (MIDI note). 
For each bar, we set the height of the matrix to 48 (considering MIDI notes from \texttt{C3} to \texttt{B6}) and the width (time resolution) to 16 (i.e., 16 time steps per bar, or equivalently 4 time steps per beat in 4/4 time signature). As a result, the size of the target output tensor for melody generation is 4 (bars) $\times$ 16 (time steps) $\times$ 48 (MIDI notes) $\times$ 1 (track).

\section{Methodology}
\label{section:method}

\subsection{Model Architecture}

Following \cite{liu18ismirlbd}, we adopt a recurrent VAE model here. In the encoder part, four-bar melody sequences are fed into bidirectional gated recurrent units (BGRU) to learn the correlation between bars. The outputs of all GRU time steps are then concatenated and passed through several dense (i.e., fully-connected) layers to get the embedding vector. In other words, given an observed input melody $\mathbf{x}$, the encoder $\mathbb{E}_{\theta}$ with parameter set $\theta$ encodes $\mathbf{x}$ into a latent vector $\mathbf{z} = \mathbb{E}_{\theta}(\mathbf{x})$.

In the decoder part, a latent vector $\mathbf{z}$ is sampled from a normal distribution characterized by $\mu$ and $\sigma$, and then passed through several fully-connected layers parameterized by $\phi$ to separately form the initial states of melody. The outputs are processed by a unidirectional GRU with a sigmoid activation layer to finally output an four-bar pianoroll. This model is illustrated in Fig. \ref{fig:PP}(a), either the upper or lower panel.

\subsection{Method 1: Fine-tuning}
As with the basic process in transfer learning, we can train the model in two stages, as illustrated in Fig. \ref{fig:PP}(b). First, we pre-train the recurrent VAE model with TT. In this stage, the goal of training is to let the model learn \emph{what melody is}. 
Here, the learning rate is set to $1e^{-3}$, and the ADAM optimization algorithm is used to accelerate stochastic gradient descent. 
As for the objective function, we use the classic binary cross-entropy and Kullback–Leibler divergence (KLD) losses. The model parameters are obtained by maximizing the following variational lower bound:
\begin{equation}
\mathcal L(\theta ,\phi; \mathbf{x}) = \mathcal L_\text{recon}(\mathbf{x}) + \mathcal L_\text{lat}(\mathbf{x}) \,, 
 \label{eq:1}
\end{equation}
where  $\mathcal L_\text{recon}=\mathbb{E}_{q_{\phi}(\mathbf{z}\mid \mathbf{x})}[\log{p_{\theta }(\mathbf{x}\mid \mathbf{z})}]$ is the reconstruction term, and $\mathcal L_\text{lat} = -D_{KL}(q_{\phi}(\mathbf{z}\mid \mathbf{x})\parallel p(\mathbf{z}))$ regularizes the encoder to align the approximate posterior $q_{\phi}(\mathbf{z}\mid\mathbf{x})$ with the prior distribution $p(\mathbf{z})$. 
$p_{\theta}(\mathbf{x}\mid\mathbf{z})$ is the data likelihood.

In the second stage, we fine-tune the model using the Jazz dataset CY$+$R. The goal is to let the model learn \emph{what Jazz is}.
The learning rate in this stage is set to $1e^{-5}$ for $t<40$, $1e^{-7}$ for $40\leq	t <80$, and $1e^{-9}$ for $t\geq80$, where $t$ denotes the epoch. Other parameters remain the same.

\subsection{Method 2: Multitask Learning}
Here, we additionally concatenate a one-hot genre label $y$ to the latent vector $\mathbf{z}$, as shown in Fig. \ref{fig:PP}(c). We train the model on both TT and CY$+$R  at the same time and regard them as two different tasks for the model to work on. For the Jazz dataset CY$+$R, we set the label $y = [0,1]$; for TT, we set the label $y = [1,0]$.

To evaluate whether our model learns to generate two different types of melodies, we pre-train a genre classifier $C(\cdot)$ to see if the output melody has the Jazz elements. The classifier basically has the same structure of the VAE encoder, but changes the output size of the final fully-connected layer to 1. When a generated melody passes through the genre classifier, the classifier outputs the probability of Jazz. We apply sigmoid activation to the output neuron of the last layer, and optimize the classifier using a cross-entropy loss. The training goal is to output 1 for Jazz and 0 for non-Jazz.
As a result, after the output melody is generated by our VAE model, it will be passed through the classifier and a probability $\hat{y} = C(\mathbf{x})$ will be obtained. 


We define the genre prediction loss as $L_\text{genre} (\hat{y},y)$ and add it to the objective function of VAE. Accordingly, the recurrent VAE model trained under the multitask learning based transfer learning method is optimized with the following objective function:
\begin{equation}
  \begin{aligned}
  \mathcal L(\theta ,\phi ;\mathbf{x},y) =~& \mathcal L_\text{recon}(\mathbf{x},y)
     + \mathcal L_\text{lat}(\mathbf{x}) + L_\text{genre} (\hat{y},y) \, 
 \end{aligned}
 \label{eq:2}
\end{equation}
Please note that, unlike the case in Eq. (\ref{eq:1}), here the reconstruction loss $L_\text{recon}$ additionally consider the provided genre label $y$.

\section{Evaluation Setup}
\label{section:exp}

\subsection{Evaluated Models}
\label{section:exp_baseline}

As our goal is to compare the effectiveness of the fine-tuning method and the multi-task learning method, we implement both of them in the evaluation. In order to examine how the transfer learning methods benefit from increasing the size of the source domain training data, we consider six levels of source-to-target data ratios ($R$):
\begin{equation}
 \begin{aligned}
R = \frac{\text{number of non-Jazz training phrases}}{\text{number of Jazz training phrases}},
 \end{aligned}
\end{equation}
where $R \in \{1,2,...,6\}$.
Moreover, as depicted in Fig. \ref{fig:PP}(a), we implement two baseline models. The first model is trained on the small, Jazz-only CY+R dataset; this can be considered as the case when $R = 0$. This model may not even learn \emph{what melody is} because the training set is really small.
The second baseline model is trained only on the large, genre-unspecified TT dataset; this can be considered as the case when $R = \infty$.
This model may not learn \emph{what Jazz is}  because the training set contains music of arbitrary genres.

\subsection{Feature Metrics}
\label{section:exp_metrics}
In order to evaluate the quality of generated melody, some features are extracted based on the work of Yang and Lerch \cite{yanglerch}. The features describe two aspects of music, including pitch- and rhythm-related ones.

The pitch-related features, including the following four, describe the preferences for arranging pitch:
\begin{itemize}
    \item \textbf{Pitch count (PC)}: The pitch count is the number of unique pitches within a phrase. The output is a scalar for each phrase.
    \item \textbf{Pitch class histogram (PCH)}: The pitch class histogram is a 12-dimensional, octave-independent representation of the pitch content for achromatic scale \cite{pcp}. 
    \item \textbf{Pitch class transition matrix (PCTM)}: The transition of pitch classes contains useful information for tasks such as key detection, chord recognition, and genre recognition \cite{yanglerch}. The two-dimensional pitch class transition matrix is a histogram-like representation computed by counting the pitch transitions for each (ordered) pair of notes. The resulting matrix size is 12$\times$12.
    \item \textbf{Pitch range (PR)}: The pitch range is calculated as the difference between the highest and lowest MIDI pitches in semitones within a phrase. The output is a scalar for each phrase.
\end{itemize}

The rhythm-related features, encompassing the following three, describe how the notes are arranged:
\begin{itemize}
    \item \textbf{Note count (NC)}: The note count is the number of notes within a phrase. 
    As opposed to the pitch count, the note count does not contain pitch information, but a rhythm-related feature that records only how many notes are in the phrase. 
    The output is a scalar for each phrase.
    \item \textbf{Note length histogram (NLH)}: To extract the note length histogram, we define a set of allowable beat length classes [full, half, quarter, 8th, 16th, dot half, dot quarter, dot 8th, dot 16th, half note triplet, quarter note triplet, 8th note triplet]. The length of a bar is defined to contain 96 unit lengths, and each note length is quantized to the nearest number of unit lengths. The rest option, when activated, will double the vector size to represent the same length classes for rests. The output vector has a length of either 12 (for notes) or 24 (12 for notes and 12 for rests), respectively.
    \item \textbf{Note length transition matrix (NLTM)}: Similar to PCTM, the note length transition matrix provides useful information for rhythm description. 
    The matrix size is 12$\times$12 or 24$\times$24.
\end{itemize}

\subsection{Overlapping Area (OA)}
\label{section:exp_oa}
Yang and Lerch [17] proposed to use Overlapping Area (OA) as an evaluation measure. The rational is given below. To compare different output sets, relative measurements may be a better choice instead of using the mean of features directly. Through relative measurements, the diversity of the dataset can be obtained. 

There are three steps in calculating the OA:
\begin{enumerate}[1.]
\item A pairwise exhaustive cross-validation is first performed for each feature. In each cross-validation step, the Euclidean distance of one sample to each of the other samples is computed. If the cross-validation is conducted on the samples in the same set, the intra-set distances are calculated. If we compare each sample in one set with all samples in another set, we calculate the inter-set distances. The output of the cross-validation process is a histogram of distances for each feature. 
    
\item In order to smooth the histogram results for a more general representation, kernel density estimation \cite{Silverman} is applied to convert the histogram into a Probability
Distribution Function (PDF).
    
\item After getting the PDFs of the target dataset and the generated melodies, OA is used to compare them. Since the melodies are generated under random sampling conditions of a Gaussian distribution, there is no overfitting problem.
\end{enumerate}

The Kullback-Leibler Divergence (KLD) is commonly used to compare two distributions. However, since in discrete probability distributions, KLD is calculated in an element-wise manner, PDFs with an identical shape (as indicated by similar Kurtosis and Skewness) but shifted on the x-axis (distinct in the mean value) yield insignificant differences in KLD. In this case, OA is able to indicate the differences. 

\section{Experimental Results}
\label{section:result}


\begin{table}[!t]
\caption{Overlapping area (OA) between the training melodies and the melodies generated by method 1.}
\begin{tabular}{|l|c|c|c|c|c|c|}
\hline
\multirow{2}{*}{}             & \multicolumn{6}{c|}{Method 1: fine-tuning}                                     \\ \cline{2-7} 
                              & R = 1  & R = 2           & R = 3           & R = 4  & R = 5  & R = 6           \\ \hline
NC                            & 0.7997 & 0.7941          & \textbf{0.8385} & 0.7690 & 0.7713 & 0.7677          \\
NC/bar                        & 0.8006 & \textbf{0.8380} & 0.8280          & 0.8005 & 0.7939 & 0.7963          \\
NLH                           & 0.7286 & 0.7185          & \textbf{0.7485} & 0.7203 & 0.7074 & 0.6976          \\
NLTM                          & 0.9158 & 0.8850          & \textbf{0.9179} & 0.8855 & 0.8869 & 0.8856          \\ \hline
PC                            & 0.6387 & 0.5986          & \textbf{0.6859} & 0.6515 & 0.6690 & 0.6770          \\
PC/bar                        & 0.8106 & 0.8123          & \textbf{0.8510} & 0.7833 & 0.7919 & 0.7851          \\
PR                            & 0.6676 & 0.6436          & \textbf{0.7134} & 0.6956 & 0.6824 & 0.7063          \\
PCH                           & 0.3198 & 0.3029          & \textbf{0.3733} & 0.3424 & 0.3679 & 0.3545          \\
PCTM                          & 0.6091 & 0.6227          & 0.6113          & 0.6918 & 0.7120 & \textbf{0.7355} \\ \hline
\multicolumn{1}{|c|}{average} & 0.6990 & 0.6906          & \textbf{0.7298} & 0.7044 & 0.7092 & 0.7117          \\ \hline
\end{tabular}
\label{table:method_1}
\end{table}

\begin{table}[!t]
\caption{Overlapping area (OA) between the training melodies and the melodies generated by method 2.}
\begin{tabular}{|l|c|c|c|c|c|c|}
\hline
\multirow{2}{*}{} & \multicolumn{6}{c|}{Method 2 : multitask learning}                                               \\ \cline{2-7} 
                  & R = 1           & R = 2           & R = 3           & R = 4           & R = 5  & R = 6           \\ \hline
NC                & 0.8536          & \textbf{0.8724} & 0.8412          & 0.7972          & 0.8002 & 0.8098          \\
NC/bar            & 0.8449          & 0.8474          & \textbf{0.8562} & 0.8073          & 0.8218 & 0.8278          \\
NLH               & 0.7661          & \textbf{0.7713} & 0.7631          & 0.7485          & 0.7341 & 0.7621          \\
NLTM              & \textbf{0.9292} & 0.9166          & 0.9132          & 0.9055          & 0.8874 & 0.8631          \\ \hline
PC                & 0.7157          & \textbf{0.7434} & 0.7349          & 0.7344          & 0.7146 & 0.7219          \\
PC/bar            & 0.8582          & 0.8593          & \textbf{0.8637} & 0.8176          & 0.8157 & 0.8184          \\
PR                & 0.7544          & 0.7261          & 0.7470          & 0.7299          & 0.7320 & \textbf{0.7585} \\
PCH               & 0.3938          & 0.3862          & 0.3478          & \textbf{0.4062} & 0.2810 & 0.3991          \\
PCTM              & 0.6670          & 0.6290          & 0.6575          & 0.6982          & 0.7142 & \textbf{0.7416} \\ \hline
average           & \textbf{0.7536} & 0.7502          & 0.7472          & 0.7383          & 0.7223 & 0.7447          \\ \hline
\end{tabular}
\label{table:method_2}
\end{table}

\begin{table}[!t]
\caption{The performances of different methods. }
\centering
\begin{tabular}{|l|c|c|c|c|}
\hline
        & Baseline 1      & Baseline 2      & Method 1 & Method 2   \\
        & (source)        & (target)        & (R=3)    & (R=1) \\ \hline
NC      & 0.7847          & 0.8287          & 0.8385   & \textbf{0.8536} \\
NC/bar  & 0.8200          & 0.8393          & 0.8280   & \textbf{0.8449} \\
NLH     & 0.6914          & 0.7407          & 0.7485   & \textbf{0.7661} \\
NLTM    & 0.8520          & 0.9081          & 0.9179   & \textbf{0.9292} \\ \hline
PC      & 0.6530          & 0.7039          & 0.6859   & \textbf{0.7157} \\
PC/bar  & 0.8020          & 0.8544          & 0.8510   & \textbf{0.8582} \\
PR      & 0.7455          & 0.7217          & 0.7134   & \textbf{0.7544} \\
PCH     & 0.3997          & \textbf{0.4531} & 0.3733   & 0.3938          \\
PCTM    & \textbf{0.7432} & 0.6909          & 0.6113   & 0.6670          \\ \hline
average & 0.7213          & 0.7490          & 0.7298   & \textbf{0.7536} \\ \hline
\end{tabular}
\label{table:baseline}
\end{table}

\begin{figure*}[!t]
\centering
\includegraphics[width=0.65\columnwidth]{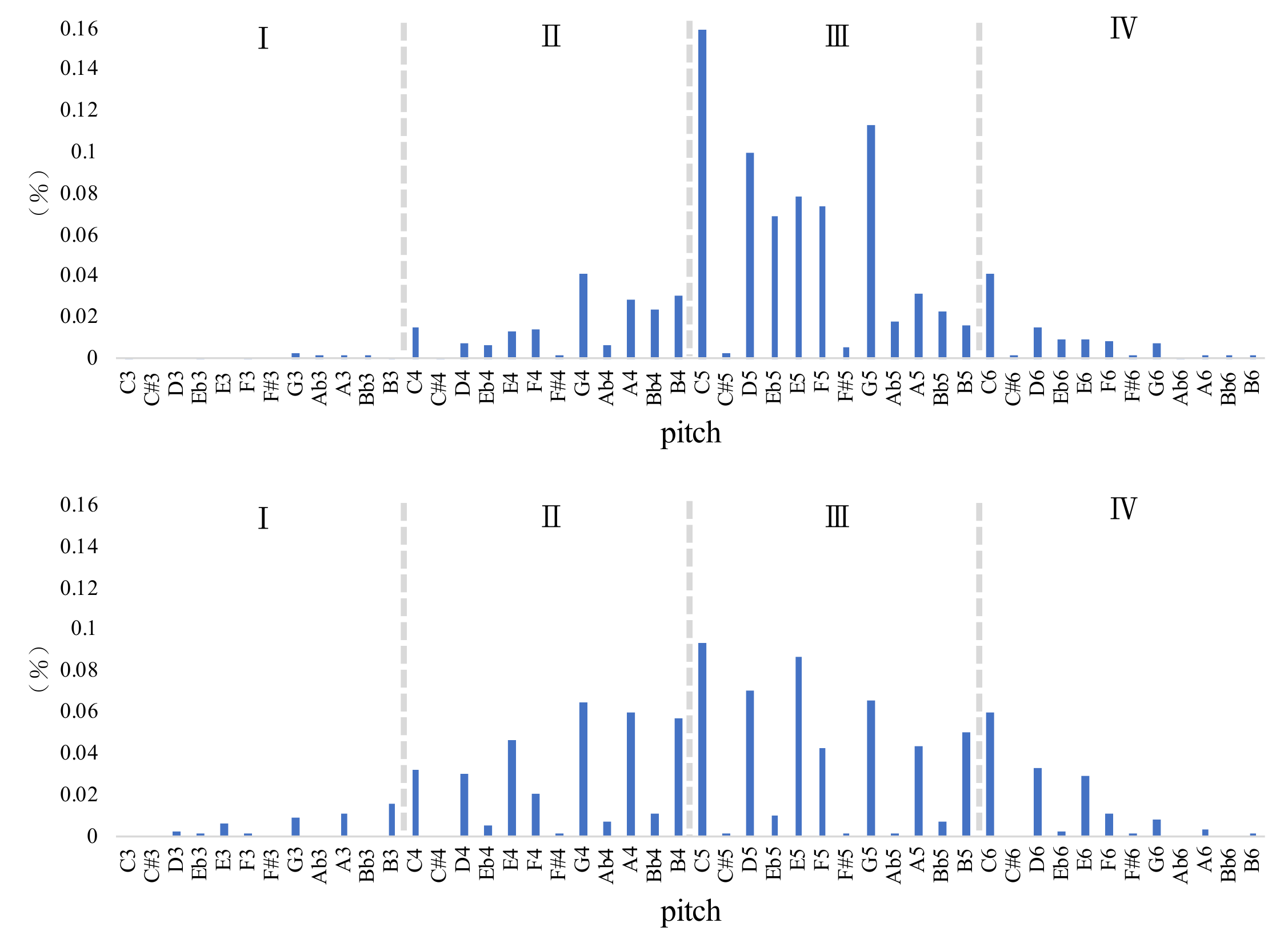}~~
\hspace{0.5cm}
\includegraphics[width=0.65\columnwidth]{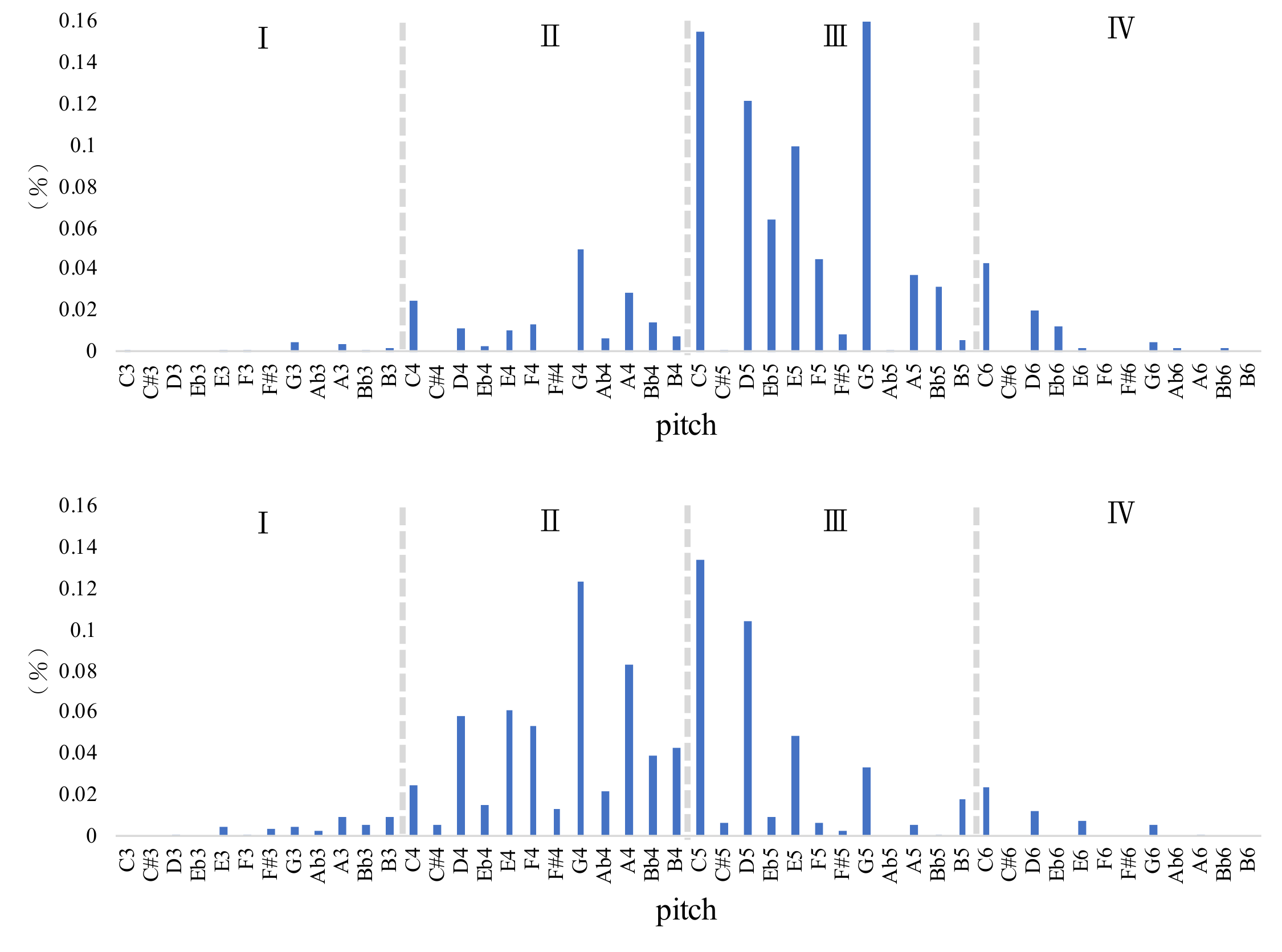}~~
\hspace{0.5cm}
\includegraphics[width=0.65\columnwidth]{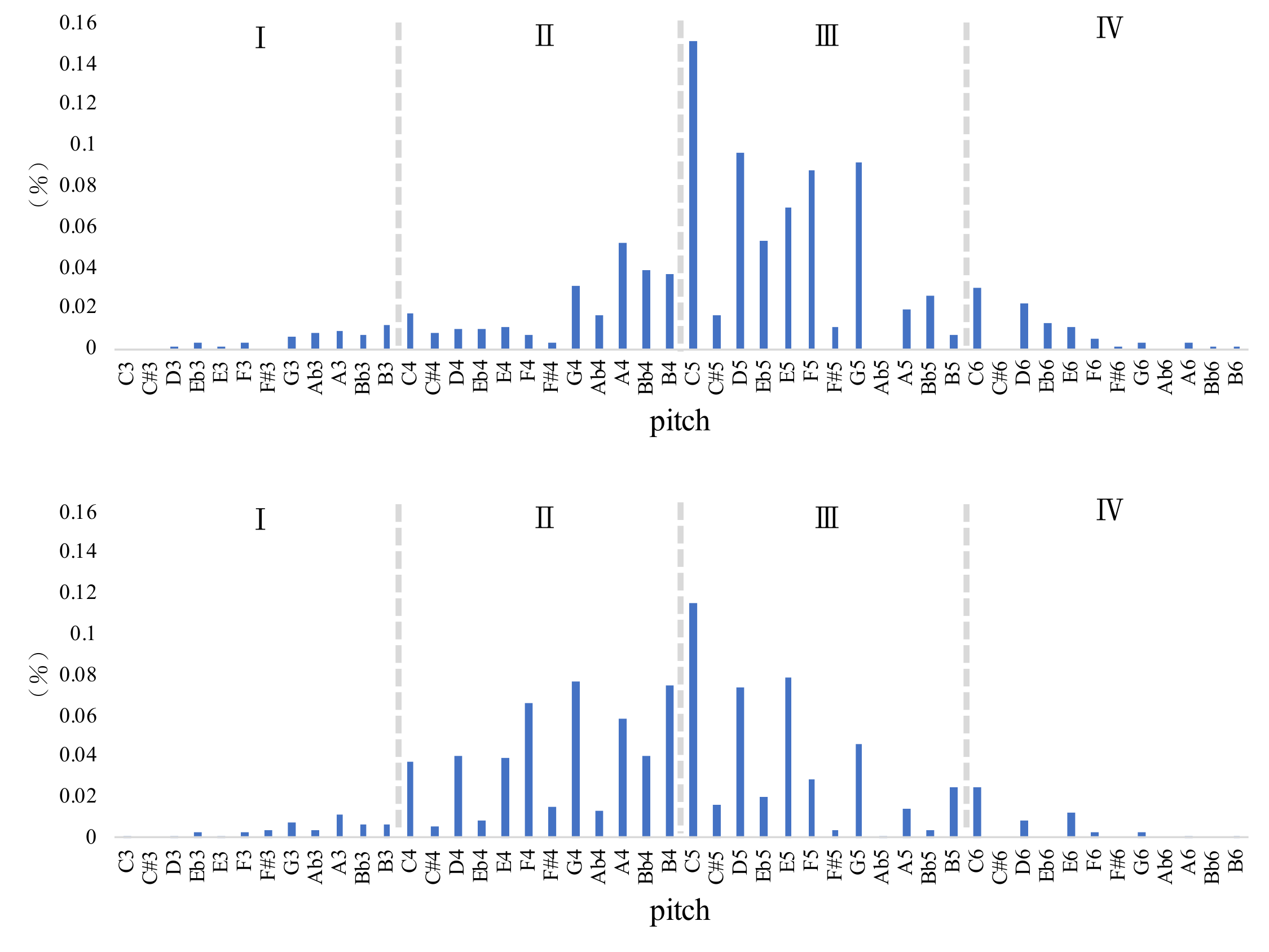}
\\ 
\hspace{0.5cm}(a)\hspace{6cm}(b)\hspace{6cm}(c)
\caption{Pitch histograms of \texttt{(a)} the source (TT) dataset (upper) and the target (CY+R) dataset (lower),  \texttt{(b)} the melodies generated by Baseline 1 trained on TT (upper)and Baseline 2 trained on CY+R (lower), and \texttt{(c)} the melodies generated by Method 1 with $R=3$ (upper) and Method 2 with $R=1$ (lower).}
\label{fig:bph}
\end{figure*}

\begin{figure*}[!t]
\centering
\includegraphics[width=0.6\columnwidth]{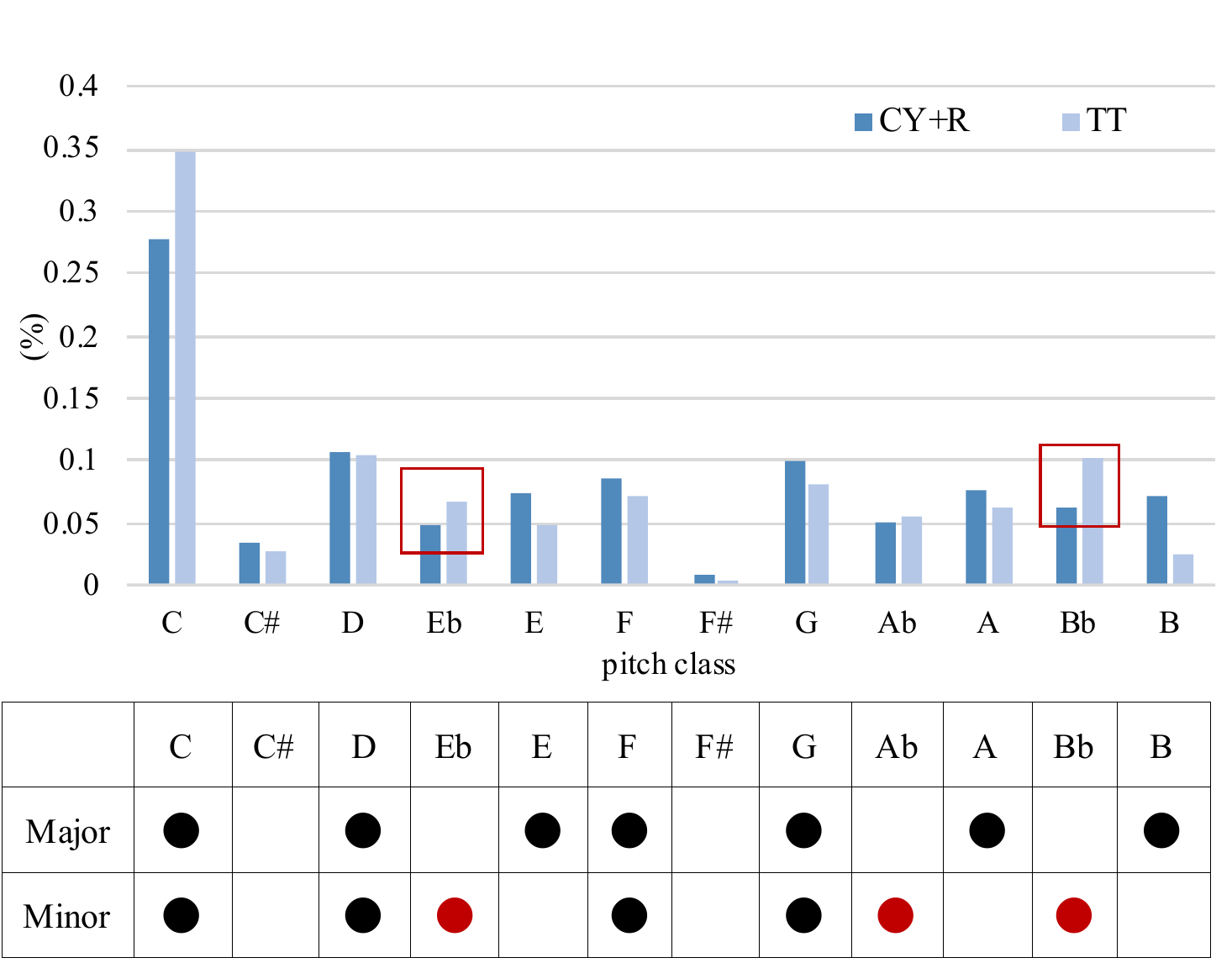}~~
\hspace{1cm}
\includegraphics[width=0.8\columnwidth]{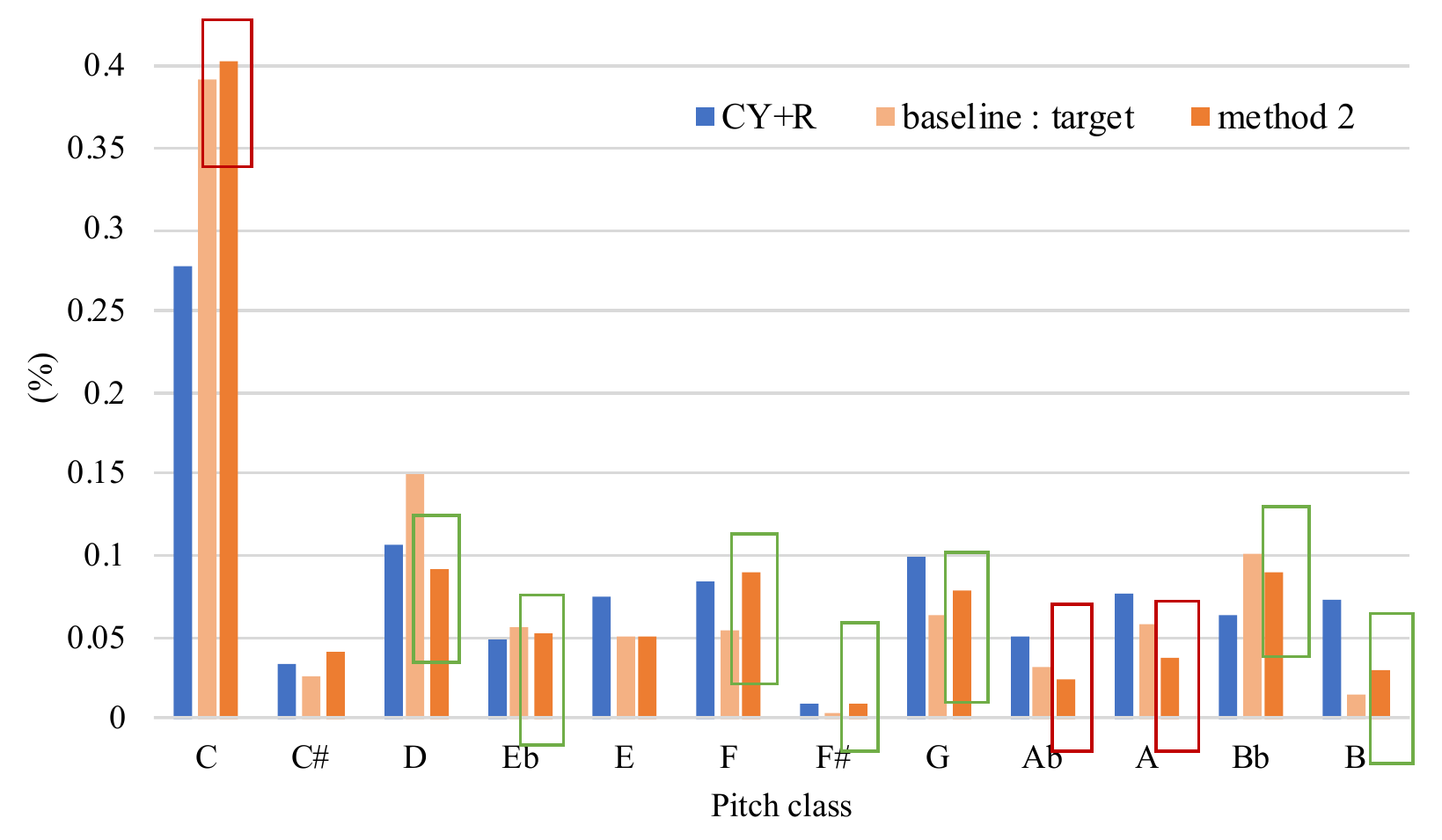}
\\ 
\hspace{0.5cm}(a)\hspace{6cm}(b)
\caption{\texttt{(a)} Upper: Pitch class histograms of the target (CY+R) and source (TT) datasets. Lower: the scales of C-Major and C-Minor; \texttt{(b)} Pitch class histograms of the target dataset and the melodies generated by Baseline 2 and Method 2 ($R=1$). }
\label{fig:pch}
\end{figure*}

\subsection{Overlapping Area}
\label{subsection:oa}
Tables \ref{table:method_1} and \ref{table:method_2} show the OAs of two transfer learning methods evaluated on different features under six levels of source-to-target data ratios (R). The bold number indicates the highest OA for each feature under different Rs. For example, the highest OA of NC for Method 1 (fine-tuning) is 0.8385 when $R=3$. 

From Table \ref{table:method_1}, we observe that $R=3$ gives the best performance in most features and the best average performance for Method 1. In contrast, in Table \ref{table:method_2}, although the best performances for different features are quite divergent between different Rs, Method 2 (multitask learning) seems to perform better when R is smaller. The reason may be due to the imbalance of source and target training data. Method 2 is more affected by the imbalance of source and target training data because its model is trained on both the source and target training datasets at the same time. Overall, the results in Tables \ref{table:method_1} and \ref{table:method_2} indicate that Method 2 (multitask learning) outperforms Method 1 (fine-tuning).

In Table \ref{table:baseline}, we compare the performances of different methods, including Baseline 1 (the model is trained on the source training dataset), Baseline 2 (the model is trained on the small target training dataset), Method 1 (fine-tuning with $R=3$), and Method 2 (multitask learning with $R=1$). Several observations can be drawn from the table. First, Baseline 2, in which the model is trained on the small target dataset, outperforms Baseline 1 whose model is trained on the large source dataset. Second, Method 1 can improve the model trained on the source dataset by fine-tuning it with a small target dataset, but the improvement is not much. Third, surprisingly, Method 1 is worse than Baseline 2, indicating that instead of fine-tuning a model trained on the source dataset with the target dataset, it is better to directly train the model on the target dataset. This result is clearly not in line with expectations and deserves in-depth study. Fourth, Method 2 is superior to the other three methods in most features except PCH and PCTM.

\subsection{Pitch-related Analysis}
In order to take a closer look to how these models manage the pitches, basic pitch histogram and pitch class histogram are drawn as Figs. \ref{fig:bph} and \ref{fig:pch}, respectively. 

Basic pitch histogram indicates the probability of presence of every pitch from C3 to B6. The pitch range covers four octaves, denoted as \uppercase\expandafter{\romannumeral 1}, \uppercase\expandafter{\romannumeral 2}, \uppercase\expandafter{\romannumeral 3}, and \uppercase\expandafter{\romannumeral 4}, in Fig. \ref{fig:bph}. In Fig. \ref{fig:bph}(a), we can see that most of the pitches in the target dataset fall in octaves \uppercase\expandafter{\romannumeral 2} and \uppercase\expandafter{\romannumeral 3}, while most of the pitches in the source dataset fall in octave \uppercase\expandafter{\romannumeral 3}. In other words, Jazz music has a wider range of pitches than generic music. In Fig. \ref{fig:bph}(b), the melodies generated by baseline methods seem to lose the diversity of pitch, in particular for Baseline 1, the pitches of the generated melodies tend to accumulate in octave \uppercase\expandafter{\romannumeral 3}. For Baseline 2, although there are quite a few pitches of the generated melodies in octave \uppercase\expandafter{\romannumeral 3}, there are much more pitches in octave \uppercase\expandafter{\romannumeral 2}. In Fig. \ref{fig:bph}(a), it is obvious that the basic pitch histogram of Method 2 is more similar to that of the target training data than Method 1. In summary, the basic pitch histogram of Method 2 is most similar to that of the target training data.

Pitch class histogram shows how the scales are used in melodies regardless of octaves. The lower part of Fig. \ref{fig:pch}(a) shows the scales of C-Major and C-Minor. As mentioned in Sec. \ref{section:db}, the TT dataset contains both C-Major and C-Minor scales. As a result, TT has higher probabilities in Eb and Bb, as highlighted with the red box in Fig. \ref{fig:pch}(a). Fig. \ref{fig:pch}(b) compares the pitch class histograms of the target CY+R dataset and the melodies generate by Baseline 2 and Method 2. Although Method 2 achieves a lower OA in PCH as shown in Table \ref{table:baseline}, we can see in Fig. \ref{fig:pch}(b) that Method 2 actually performs better than Baseline 2 in many scales highlighted with the green box. A possible reason is that the OA of PCH in Table \ref{table:baseline} is calculated in a sample-wise manner, but the PCH in Fig. \ref{fig:pch}(b) presents the overall pitch class distributions of the target dataset and the melodies generated by the models. This means that Method 2 learns the probability of the presence of a scale in Jazz music, but does not mean that the combination of pitches within each sample (i.e., a four-bar melody phrase) generated by it conforms to the overall target distribution.

\subsection{Subjective Test}
\label{subsection:test}

\begin{table}[!t]
\caption{The results of the subjective test.}
\label{table:sub}
\begin{tabular}{|c|c|cccc|}
\hline
       & CY+R   & \begin{tabular}[c]{@{}c@{}}Baseline 1\\ (source)\end{tabular} & \begin{tabular}[c]{@{}c@{}}Baseline 2\\ (target)\end{tabular} & \begin{tabular}[c]{@{}c@{}}Method 1\\ (R=3)\end{tabular} & \begin{tabular}[c]{@{}c@{}}Method 2\\ (R=1)\end{tabular} \\ \hline
Type \uppercase\expandafter{\romannumeral 1} & 3.5905 & 2.6286                                                        & 2.7524                                                        & 2.8095                                                   & \textbf{2.8857}                                          \\
Type \uppercase\expandafter{\romannumeral 2} & 3.7744 & 2.3179                                                        & 2.3282                                                        & \textbf{2.6564}                                          & 2.3538                                                   \\
Type \uppercase\expandafter{\romannumeral 3} & 3.8500 & 2.2750                                                        & 2.4000                                                        & \textbf{2.8500}                                          & 2.5250                                                   \\ \hline
\end{tabular}
\end{table}

\begin{figure*}[!t]
\centering
\includegraphics[width=2\columnwidth]{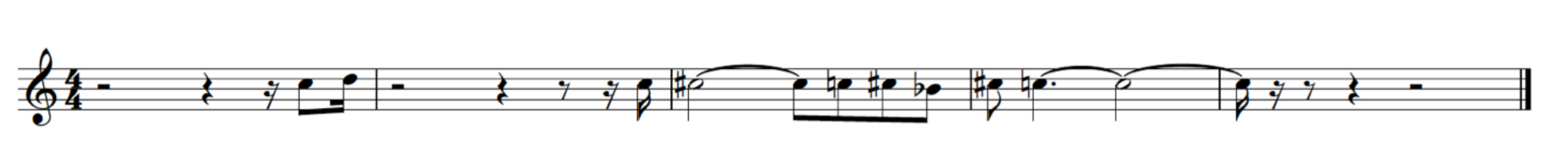}\\
\hspace{0cm}(a)\\
\includegraphics[width=2\columnwidth]{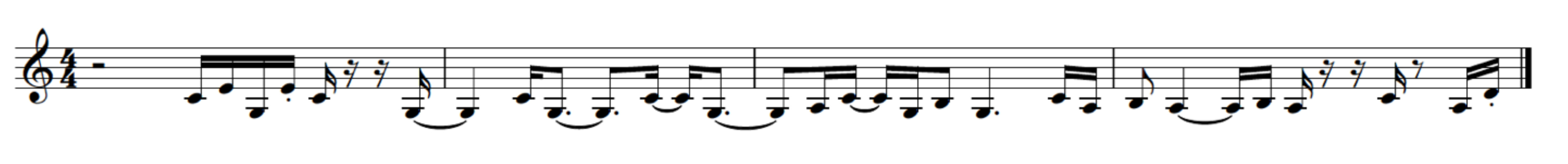}\\
\hspace{0cm}(b)
\caption{Score of melody generated by \texttt{(a)} method 1, \texttt{(b)}method 2. }
\label{fig:compare}
\end{figure*}

In addition to the objective test, we also conducted a subjective listening evaluation. We let the subjects listen to two demo melodies from the CY+R dataset, and asked them to rate five groups of four-bar melody phrases. Each group contained five melody phrases, one from the CY+R dataset, and the remaining four were generated by two baselines, Method 1 with $R=3$, and Method 2 with $R=1$, respectively. After listening to each test melody, the subjects were asked to give a score in a five-point Likert scale according to the degree of similarity between the test melody and the demo melody. The subjects were also required to provide information about their musical expertise. They could choose the category that best fits them from\\ 
\begin{itemize}
\item[] Type \uppercase\expandafter{\romannumeral 1}: seldom listening to soft jazz,
\item[] Type \uppercase\expandafter{\romannumeral 2}: a music lover, and listening to jazz (soft jazz) sometimes, and
\item[] Type \uppercase\expandafter{\romannumeral 3}: professional composer.
\end{itemize}

The results of the subjective test are shown in Table \ref{table:sub}. 69 subjects participated in the test, of which 21 belonged to Type \uppercase\expandafter{\romannumeral 1}, 39 belonged to Type \uppercase\expandafter{\romannumeral 2}, and 9 belonged to Type \uppercase\expandafter{\romannumeral 3}. From the results, the following observations can be drawn.
\begin{itemize}
    \item Both Method 1 and Method 2 scored higher than the two baselines. 
    \item Method 2 got the highest score for Type \uppercase\expandafter{\romannumeral 1} subjects.
    \item The subjects for Types \uppercase\expandafter{\romannumeral 2} and \uppercase\expandafter{\romannumeral 3} preferred the melodies generated by Method 1 instead of Method 2.
\end{itemize}

The reason for the difference between objective and subjective tests might be Type 2 and type 3 objectives are more aware of the exist of some jazz-related technique. As a result, when such pattern appears in a melody, objectives tend to consider it more like real data. For example, \ref{fig:compare} shows the score of melodies generated by method 1 and method 2 in one of the subjective test rounds. In the third bar of \ref{fig:compare}(a), there seems like an "chromatic enclosures" , which has been a part of jazz vocabulary since Bebop. As a result, (a) get a higher score of 3.44 than (b), which is 1.78. Perhaps we should have adopted an algorithm for finding such musical patterns as an evaluation metric.

\section{Conclusions}
\label{section:conclusion}
In this paper, we proposed using a recurrent VAE to randomly generate a jazz melody. We compared two methods of utilizing a big source dataset for transfer learning with a small target dataset and investigated the influence of the source-to-target data ratio. The overlapping areas computed based on the distributions of different pitch-related and rhythm-related features demonstrates that the multitask learning-based method (Method 2 in this paper) is slightly better than the fine-tuning-based method (Method 1) and the baseline methods that train the model on either the source dataset or the target dataset. On the other hand, the subjective test shows that the subjects who sometimes listen to soft jazz or are professional about composition think the melodies generated by Method 1 are better than those generated by Method 2. In our future work, we will try to figure out the reason for the difference between objective and subjective tests. 

\section*{Acknowledgment}
This work was supported in part by the Ministry of Science and Technology of Taiwan under Grant: MOST 105-2221-E001-012-MY3.

\end{document}